# A Method of Passage-Based Document Retrieval in Question Answering System

Man-Hung Jong, ,

College of Computer Science, Kim Il Sung University, Pyongyang, D.P.R.K

Chong-Han Ri

College of Computer Science, Kim Il Sung University, Pyongyang, D.P.R.K

Hyok-Chol Choe

College of Computer Science, Kim Il Sung University, Pyongyang, D.P.R.K

Chol-Jun Hwang

Electronic Library, Kim Il Sung University, Pyongyang, D.P.R.K

**Abstract:** We propose a method for using the scoring values of passages to effectively retrieve documents in a Question Answering system.

For this, we suggest evaluation function that considers proximity between each question terms in passage. And using this evaluation function , we extract a documents which involves scoring values in the highest collection, as a suitable document for question.

The proposed method is very effective in document retrieval of Korean question answering system.

## 1. Introduction

It is often important in building a Korean Question Answering (QA) system to decrease the number of documents related with a question by retrieving documents that include some information to the answer for a given question, because the performance of a QA system should be increased by searching the answer in a limited time after retrieving appropriate documents for the question. Thus, many approaches has been proposed for retrieving documents related with the question from the collection[1, 2, 3, 4].

In many document retrieval systems, the approaches has been proposed that apply the vector spac model (VSM), pseuo-feedback and latent semantic indexing(LSI) based methods often used in general information retrieval systems[1, 2, 5]. However, these approaches are not appropriate for the document retrieval in QA systems because the estimate the appropriate documents according to the number of query terms involved in a document without considering the distribution of the query terms.

In general, it has seen that smaller text area the query terms occur in, more appropriate the document is to the user's question. [6, 7]

To overcome this, we introduce a new concept: "passage" as a new retrieval units and has presented the method which retrieval the document on based score value of passage.

These approaches segment a document into passages with a given size (3 sentences or about 300 words) and then use *tf-idf* model or BM25 model to score passages according to the similarity between query and passage and to rank documents with the highest similarity. The disadvantage of them, however, is that the size of passages is fixed and so if the size is too small, the query terms with long distance would not be covered in a passage and if the size is too large the density distribution would not be reflected.

We define a scoring function of documents for a user question using the size of passages and a method using the scores of passages to rank documents in document retrieval.

## 2. Document Retrieval method using passages

Most document retrieval methods using passages segment a document into passages with fixed size and then retrieve documents according to the scores of passages in them. In some cases, the variants of this approach have been used in [5].

We propose a scoring function based the proximity of query terms considering passages with different lengths and a ranking method using it.

To summarize the document retrieval method proposed in this paper are as follows.

・First search the occurance position of every query term in document $d_j \in D$.

・Next, finde all the possible passage containing all the query terms document $d_j$.

And calculate the score of every passage by using the evaluation-function considering the proximity between the query terms.

・Output the document containing the highest score passage as the document fit to the query.

### 2.1 Selection of passage

Let the set of query terms occurring in the query be $Q$.

That is: $Q = \{q_1, q_2, ......, q_t\}$

Denote the set of possible occurrence position of query term $q_i$ in a document $d_j$ as $P_t^{d_j}$.

$$P_i^{d_j} = \{p_{i1}^{d_j}, p_{i2}^{d_j}, ......, p_{ini}^{d_j}\}$$

Where $p_{ik}^{d_j}$: $k \cdot th$ occurance position of query term $q_i$ in document $d_j$

$n_i$: size of $P_i^{d_j}$

To figure out this is as follows.

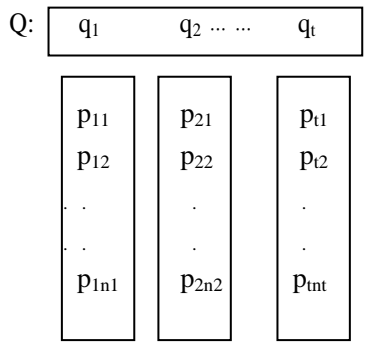

$$P_1 \quad P_2 \quad \ldots \quad P_t$$

Thus, the set of possible passages $\Omega^{d_j}$ that query term set $Q \in \{q_1, q_2, \ldots, q_t\}$ occurs in document $d_j$ is as follows.

### 2.2 Calculating the score of a passage

The score of possible passage in document $d_j$ is calculated using the evaluating function.

#### 2.2.1 Evaluating the proximity between arbitrary two query terms in a passage

Then, the proximity between terms, $q_i$ and $q_j$ in passage $\omega$ is calculated using the following function.

$$CP(p_i, p_j) = \frac{1}{1 + s \bullet \ln(1 + dist(p_i, p_j))}$$

Where $dist(p_i, p_j)$: the distance between query terms, $q_i$ and $q_j$

$p_i$: the position of term $q_i$ in passage $\omega$

$p_j$: the position of term $q_j$ in passage $\omega$

$s$: $s \in R_+$, a parameter representing the importance of the distance

This function value is 1 when two query terms are neighboured, and is converged to 0 the more distant they are.

#### 2.2.2 Evaluating the proximity of al query terms in a passage.

The proximity of all query terms in a passage is evaluated using the following function.

That is, the scores of possible passages in a document are calculated as follows.

$$R^{d_j}(\omega) = \frac{s(\omega)}{\max s(\omega')} \qquad \omega, \omega' \in \Omega^{d_j}$$

Where $s(\omega) = \sum_{i=1}^{t} \sum_{j=i+1}^{t} CP(p_i, p_j)$

### 3. Document retrieval using the score of passage

The score of document $d_j$ is ev.aluated as follows

$$DS(d_j^*, \omega^*) = \max_{d_j, \omega \in \Omega^{d_j}} R^{d_j}(\omega)$$

Where $d_j^*$: the most appropriate document to the query

$\omega^*$: the most appropriate passage in $d_j^*$

$\omega \in \Omega^{d_j}$: possible passages in document $d_j$

### 4. Experimental Results & Analysis

In order to evaluate the performance of document retrieval using the proposed scoring function, we use 360 standard queries and answering documents from "Korean Complete History" (vol. 1~5) as a test data. Then we use MRR(Mean Reciprocal Rank) as a measure that is often used for the evaluation of the document retrieval module in QA systems.

A Method of Passage-Based Document Retrieval in a Korean Question Answering System

Table. Evaluation Results for the Retrieval Performance

| Retrieval Model | MRR |
|---|---|
| Vector spase model(VSM) | 0.406 |
| Pseudo-Feedback | 0.501 |
| latent semantic indexing(LSI) | 0.503 |
| Retrieval model by score function | 0.511 |
| Presented method | 0.523 |

**5. Conclusion**

We presented a method for using the scoring values of passages to effectively retrieve documents in a Question Answering system.

For this , we first presented a determinating method of passage size, then we suggest evaluation function that considers proximity between each question terms in passage. And using this evaluation function , we extract a documents which involves scoring values in the highest collection, as a suitable document for question.

The proposed method is very effective in document retrieval of Korean question answering system.